\documentclass[prl,twocolumn,floatfix,superscriptaddress,nofootinbib,aps]{revtex4}

\usepackage{natbib} 
\usepackage{bm}

\usepackage{graphicx}
\usepackage{amsmath}
\usepackage{amssymb}
\usepackage{hyperref}
\usepackage{verbatim}

\newcommand{\tcm}{21\,cm}

\renewcommand{\a}{\alpha}
\renewcommand{\b}{\beta}

\renewcommand{\k}{k_a}

\newcommand{\p}{\phantom{\alpha}}

\newcommand{\td}[1]{\tilde{#1}}

\newcommand{\sectionk}[1]{
    {\it #1.}---\
  }

\begin{document}

\title{Primordial gravity waves fossils and their use in
testing inflation}

\author{Kiyoshi Wesley Masui}
\email{kiyo@cita.utoronto.ca}
\affiliation{Canadian Institute for Theoretical Astrophysics}
\affiliation{Department of Physics, University of Toronto}

\author{Ue-Li Pen}
\email{pen@cita.utoronto.ca}
\affiliation{Canadian Institute for Theoretical Astrophysics}

\date{August 25, 2010}

\begin{abstract}
A new effect is described by which primordial gravity waves leave a permanent
signature in the large scale structure of the Universe.  The effect
occurs at second order in perturbation theory and is sensitive to the
order in which perturbations on different scales are generated. 
We derive general forecasts for the detectability of the effect with future
experiments, and consider observations of the pre-reionization
gas through the 21\,cm line.
It is found that the Square Kilometre Array will not 
be competitive with
current cosmic microwave background constraints on
primordial gravity waves from inflation.
However, a more futuristic experiment could, through this effect,
provide the highest ultimate sensitivity to tensor modes and
possibly even measure
the tensor spectral index. It is
thus a potentially quantitative probe of the inflationary paradigm.

\end{abstract}

\maketitle

\sectionk{Introduction}
\label{s:intro}
It has been proposed that redshifted 21\,cm radiation, from the
spin flip transition in neutral hydrogen, might be a powerful probe
of the early universe.
The era before the first luminous objects reionized the
universe--around redshift 10--contains most of the observable 
volume of the universe, and 21\,cm
radiation is the only known probe of these so called dark
ages (see \citet{astro-ph/0608032} for a review).
The density of the hydrogen could be mapped in 3D
analogous to how the cosmic microwave background (CMB) is mapped
in 2D. The wealth of obtainable statistical information 
may allow for the detection of many subtle effects which
could probe the early universe.  In particular, the primordial
gravity wave background, also referred to as tensor perturbations, 
are of considerable cosmological interest.

Inflation robustly predicts the production of tensor
perturbations with a nearly scale-free spectrum,
however, their amplitude is essentially
unconstrained theoretically. 
The amplitude of the tensor power spectrum is
quantified by $r$, the tensor to scalar ratio.
The current upper limit is $r<0.24$ at 95\% 
confidence \citep{arXiv:1001.4538}, however
upcoming CMB measurements 
will be sensitive down to $r$ of a few
percent \citep{arXiv:1003.6108}.  The current limits on $r$
correspond to characteristic primordial shear on the order of 
$10^{-5}$ per logarithmic interval of wavenumber.

Several probes of gravity waves using the pre-reionization
21\,cm signal
have been proposed.  These include polarization
\citep{astro-ph/0702600} and
redshift space distortions \citep{arXiv:0901.3655}.  
\citet{astro-ph/0301177} considered the
weak lensing signature of gravity waves and found that the signal
is sensitive to the so called metric shear.  This is closely
related to the present work.

Here we describe a mechanism by which primordial
gravitational waves may leave an imprint in the statistics
of the large scale structure (LSS) of the universe.  
This signature becomes observable
when the gravity wave enters the horizon and begins to
decay.

\sectionk{Mechanism}
\label{s:mech}
In the following, Greek indices run from 0 to 3 and lower
case Latins from 1 to 3.  Latin indices are always raised and
lowered with Kronecker deltas.  Commas denote partial derivatives,
and an over-dot ($\dot{\#}$)
represents a
derivative with respect to the cosmological conformal time.
Finally, we adopt a mostly
positive metric signature $(-1,1,1,1)$.

We start with an inflating universe with some distribution of
previously generated tensor modes that are now super horizon.
Scalar,
vector and smaller scale tensor modes may exist but their
contribution to the metric is ignored.
The line element is given by
\begin{equation}
  \label{e:mech:CFmetric}
  ds^2 = a(\eta)^2\left[ -d\eta^2 + (\delta_{ij}+h_{ij})dx^i
    dx^j\right].
\end{equation}
where $a$ is the scale factor, $\eta$ the conformal time and a
spatially flat background geometry has been assumed.
The metric perturbations
$h_{ij}$ are assumed to be transverse and traceless and thus
contain only tensor modes.  The elements of $h_{ij}$ are
also assumed to be small such that only leading order terms need be
retained.  The assumption that all tensor
modes under consideration are super horizon implies that
$k_h \ll \dot{a}/a$, where $k_h$ denotes the wave numbers of
tensor modes.  The frame in which the line element takes the
form in Eq.~\ref{e:mech:CFmetric} will hereafter be
referred to as the cosmological frame (CF).

By the equivalence principle, it is possible to perform a
coordinate transformation such that the space-time appears
locally Minkowski at a point. New coordinates are defined in
which the tensor modes are gauged away at the origin:
\begin{equation}
  \label{e:mech:LFFcoords}
  \td{x}^\a = (x^\a + \frac{1}{2}h^\a_{\p\b}x^\b),
\end{equation}
where the elements $h_{0\a}$ are taken to be zero.
The metric now takes the form (up to first order in $h_{ij}$)
\begin{equation}
  \label{e:mech:LFFmetric}
  ds^2 = a^2\left[-d\eta^2+\delta_{ij}d\td{x}^i
    d\td{x}^j
    -\td{x}^c \partial_\alpha h_{\beta c} d\td{x}^\a
    d\td{x}^\b\right].
\end{equation}
This frame will be loosely referred to as the locally
Friedmann frame
(LFF), because in these coordinates the metric is locally that
of an unperturbed
FLRW Universe.  We will give quantities in
these coordinates a tilde ($\td{\#}$) to distinguish them from
their counterparts in the CF.
It is seen from Eq.~\ref{e:mech:LFFmetric} that the local effects of
gravity waves are suppressed not only by the smallness of
$h_{ij}$ but also by $k_h/k$ where $k=L^{-1}$ and $L$ is some
length scale of interest.  This will be important in
justifying some later assumptions.  Note that for super
horizon gravity waves, temporal derivatives are much smaller
than spacial ones.

On small scales, inflation generates scalar
perturbations which are then carried to larger scales by the
expansion.  By the equivalence principle,
physical processes on small scales can not know about the long
wavelength tensor modes.  As such these small scale scalar modes 
must be uncorrelated with the long wavelength tensor modes.  
We assume statistical homogeneity and isotropy in the LFF 
as would be expected from inflation. The power spectrum of scalar
perturbations can then be written as a function of only the
magnitude of the wave number, i.e., $\td{P}(\td{k}_a)=\td{P}(\td{k})$.
This applies only within the local patch near the point where the
tensor mode was gaged away.  The average in the definition of the 
scalar power spectrum is over realizations of the scalar map,
but not the tensor map.

In the CF, the isotropy is broken.  Transforming back to
cosmological coordinates maps $\td{k}_i \to k_i - k_j h_i^{\p j}/2$.
The power spectrum becomes sheared:
\begin{equation}
  \label{e:mech:P}
  P(k_a) = \td{P}(k)-\frac{k_i k_j h^{ij}}{2k}
     \frac{d \td{P}}{dk} +
     O(\frac{k_h}{k}h_{ij})+O({h_{ij}}^{2}).
\end{equation}
If the metric perturbations are not assumed to be traceless,
the right hand side of this equation gains an additional
term proportional to this trace.
This deviation from isotropy is not
observable since any possible observation
would take place in the LFF.

It is noted that the leading order correction to CF power spectrum is
not suppressed by $k_h/k$.
It is therefore not expected that the residual terms in
the LFF metric (Eq.~\ref{e:mech:LFFmetric}) can break
isotropy to undo CF anisotropy.  However it
was the CF in which the power spectrum should be isotropic,
then there would be \emph{observable} anisotropy in the LFF.  This would
be a violation of the equivalence principle, since an experiment local
in both space and time would be able to detect the super horizon tensor
modes by measuring the power spectrum of the locally
generated scalar perturbations.

We would now like to evolve the system to some later time when
observations can be made.
Ignoring the internal dynamics of the scalar perturbations,
we solve for their evolution as if they were embedded in a
sea of test particles. This is trivial since an
object at coordinate rest in the CF will remain at rest for
any time dependence of $h_{ij}$ (this is true at all orders).
At some point well after
inflation, when the universe is in its deceleration phase,
the horizon will become larger than the length scale of the
tensor modes.  The tensor modes will then decay by
redshifting, and after some period of time the
metric perturbations $h_{ij}$ become negligible.  The
CF and LFF then become equivalent and both correspond to the
frame in which observations can be made.  The distribution of
test particles is the same as it initially was in the CF.
As such, the initially physically isotropic power spectrum now contains a
measurable local anisotropy given by Eq.~\ref{e:mech:P}.
The values of the initial
metric perturbations can be determined by measuring this
distortion at any time in the future, constituting a
fossil of the initial tensor modes.

The scalar perturbations
remain Gaussian but become non-stationary, and the trispectrum
gains the corresponding terms.
This is analogous to the apparent distortions expected in the
CMB and \tcm{} fields induced by gravitational
lensing. Similarly the bispectra of mixed scalars and tensors
were calculated in \citet{astro-ph/0210603}, employing
similar methodology to that presented here.

The effect described here is a second order perturbation
theory effect, in that it is a small effect due to tensor
modes on the already small scalar perturbations.  This
coupling occurs in the initial conditions, not between the
dynamics of the scalars and tensors.  The simple argument
presented above avoided the complication of a full second
order calculation, but it is expected that
such calculations would yield the same
results.  Specifically,  an
expression agreeing with Eq.~\ref{e:mech:P}, to relevant
order, was derived in \citet[Eq.~4.5]{arXiv:1005.1056} as part of a
longer calculation.

\sectionk{Tests of inflation}
The above arguments
relied on perturbations on large scales being generated before
perturbations on small scales.  This is the case in any
conceivable model of inflation, however it is not be the case in
all scenarios.  As an illustrative example,
in the cosmic defect scenario
perturbations are generated on small scales and then causally
transported to larger scales as the universe evolves.  It is
argued that in this scenario, tensor perturbations leave no
fossils.
A detection of primordial tensors by
another means (CMB B-modes for example) with an observed lack
of the corresponding fossils would provide a serious
challenge to inflation.

The most specific prediction of single field inflation is the power
spectrum of tensor modes, defined by
\begin{equation}
\label{e:tests:PTdef}
(2\pi)^3\delta(\k-\k')P_h(\k) \equiv
\langle h_{ij}(\k)h^{ij}(\k')\rangle.
\end{equation}
Given the amplitude of the scalar
power spectrum $A_s$, the tensor power spectrum is fixed by a single
parameter, the tensor to scalar ratio $r$.
The shape of the spectrum is then
nearly scale-free: 
\begin{equation}
  \label{e:tests:PT}
  P_h = \frac{2 \pi^2 r A_s}{k^3} \left(\frac{k}{k_0}\right)^{n_t}.
\end{equation}
We follow the WMAP conventions
for defining $P_h$, $A_s$ and $r$ \citep{arXiv:0803.0547}. 
The spectral index
fixed by the consistency relation,
$n_t = -r/8$ \citep{2000cils.book.....L}.
The pivot
scale is taken to be $k_0= 0.002 \mathrm{\,Mpc}^{-1}$ and we
assume the WMAP7 central value for $A_s$ of
$2.46 \times 10^{-9}$.

Because $r$ is likely small, any deviation
from a scale-free spectrum will be difficult to measure,
making the verification of the consistency relation
correspondingly difficult.  The CMB is sensitive primarily to
large scale tensor modes, with smaller scale modes having
decayed by recombination.  Cosmic variance and lensing
contamination will
likely prevent a measurement of $n_t$ from the CMB, unless
the lensing can be cleaned from the
signal \citep{arXiv:0902.1851}.
Conversely, the amplitude of the fossil signal does not decay
as the universe expands.  It may thus be possible to make a
measurement of the spectral index, provided $r$ is
sufficiently large.

\sectionk{Statistical detection in LSS}
In practice, the tensor gravity wave fossils could be
reconstructed by applying quadratic estimators to the
density field.  Aside from the increased dimensionality,
this is identical to the manner in which lensing shear is
reconstructed
\citep{astro-ph/9810257,arXiv:0710.1108}. Rather than considering
the statistics of such estimators, here we follow a simpler line of
reasoning to approximate the accuracy to which the tensor
parameter can be measured.

We begin by asking how well a long wavelength,
tensor mode can be reconstructed
from its effects on the scalar power spectrum
(Eq.~\ref{e:mech:P}).  The metric perturbations are assumed
to be spatially constant and take the form 
\begin{equation}
h_{ij}=h_+e^+_{ij}(\hat{z}) + h_\times
e^\times_{ij}(\hat{z})
\end{equation}
where $e^+_{ij}$ and $e^\times_{ij}$ are the polarization
tensors and the $\hat{z}$ direction of propagation is chosen for 
convenience.
The uncertainty on the scalar power spectrum is
\begin{equation}
\label{e:fore:deltaP}
  \left[\Delta P(\k)\right]^2 = 2 \left[P(\k)+N\right]^2,
\end{equation}
where $N$ is the noise power. We use a Fisher Matrix analysis to sum
this information over all $\k$ to determine the corresponding
uncertainty on the shear $h_+$ and $h_\times$.  Assuming an
experiment whose noise is sub dominant to sample variance ($N \ll P$),
the resulting
variance is inversely proportional to the number of modes
surveyed:
\begin{equation}
\left(\Delta h^C\right)^2 \sim \left[V(k_{max}/2\pi)^3\right]^{-1},
\end{equation}
where $h$ stands
for either $h_+$ or $h_\times$ (the superscript $C$
indicates that the formula applies for spatially constant
$h$), $V$ is the volume of the survey
and $k_{max}$ is set by the resolution of the survey.   The
constant of proportionality depends on the shape of the
unsheared power spectrum $\td{P}(k)$, but to within a few tens
of percent it is unity. \tcm{}
emission will be difficult to observe on large scales
\citep{astro-ph/0608032}, however it is small scales that
dominate the number of modes and thus
the reconstruction.  It is only the coherence of
small scale anisotropy that must be measured on large
scales.

Given the reconstruction uncertainty on a spatially constant shear, and
the fact that reconstruction noise is scale independent (white)
\citep{astro-ph/9810257},
the noise power spectrum for spatially varying tensor modes 
is then
\begin{equation}
  N_h = 4V (\Delta h^{C})^2 =
  4\left(\frac{2\pi}{k_{max}}\right)^3.
\end{equation}
The factor of four comes from the definition of the
power spectrum in Eq.~\ref{e:tests:PTdef}, noting that
$\langle h_{ij}h^{ij} \rangle =4 \langle h^2 \rangle$. 

We now sum over 
$\k$\footnote{From this point forward, $\k$ will
refer to the wave number of a tensor mode, not a scalar mode.
The exception will be $k_{max}$ which is the smallest scale at
which a scalar can be resolved.} to
determine the signal to noise as a function of tensor power spectrum
amplitude $r$.
The signal to noise ratio squared is then
\begin{align}
\textit{SNR}^2 &=  \sum_{\k, \{+,\times\}} \frac{P_h^2}{2(N_h+P_h)^2}\\
      &\approx V\int_{k_{lower}}^{k_{upper}} \frac{dk\,k^2}{2\pi^2 }
        \frac{P_h^2(k)}{(N_h+P_h)^2}.
\end{align}
It is seen from the redness of the spectrum $P_h$ 
(Eq.~\ref{e:tests:PT}) that the result is
completely independent of the upper limit of integration.  The same redness
makes the final result extremely sensitive to the lower limit.
As described above, the fossil of a
primordial tensor mode can only be observed once the mode has
decayed.  This begins to happen when the scale of the gravity
wave becomes comparable to the horizon scale, and as such, the
largest scale observable mode has wavelength $k_{lower}\approx aH$.

For an initial detection, we assume that noise dominates 
sample variance at each $\k$, i.e., $N_h\gg P_h$.
Setting the signal to noise ratio to be 2, for a 95\%
confidence detection, yields a minimum detectable amplitude of
\begin{equation}
r_{min} = \frac{32 \pi^2}{A_s k_{max}^3}\left(\frac{6}{VV_H(z)}\right)^{1/2}
\end{equation}
where $V_H\equiv (aH)^{-3}$.

While the observability of \tcm{} radiation depends on the
reionizaton model, one regime in which a strong
signal may exist is near redshift 15 \citep{astro-ph/0608032}.
The planned Square Kilometer Array (SKA) will aim to probe this era
with 10\,km baselines \citep{2009IEEEP..97.1482D}.  Assuming a
survey volume of $200\,(\textrm{Gpc}/h)^3$ and a noiseless
measurement, the limit on $r$
achievable with SKA will be
\begin{equation}
r_{min} \approx 7.3\left(\frac{1.2\,\textrm{Mpc}/h}{k_{max}}\right)^3
    \left[\frac{200\,(\textrm{Gpc}/h)^3}{V}
    \frac{3.3\,(\textrm{Gpc}/h)^3}{V_H}\right]^{1/2}.
\end{equation}
While this constraint is not competitive with current
constraints from the CMB, it is a strong function of
the resolution of the experiment.  The Low Frequency Array (LOFAR) for
instance, has baselines
extending to 400\,km. However LOFAR will not have the sensitivity
to probe the dark ages \citep{astro-ph/0307240}.
It is the physical shear due to gravity waves at the source that is
being measured, and all light propagation effects, such as the lensing
considered in \citet{astro-ph/0301177}, have been ignored.

\begin{figure}
  \includegraphics[scale=1]{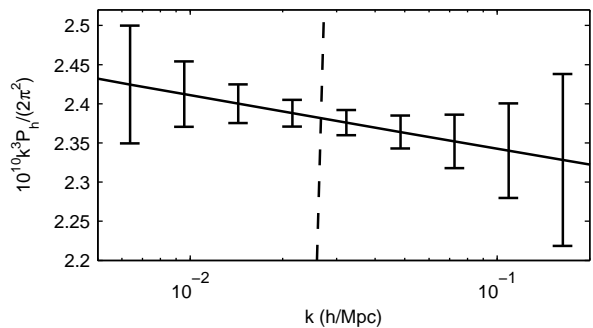}
  \caption{\label{f:tensorpower}  Primordial tensor power spectrum
  obeying the consistency relation for $r=0.1$. The solid line is the
  tensor power spectrum.  Error bars represent the
  reconstruction uncertainty on the binned
  power spectrum for a perfect experiment,
  surveying $200\,(\textrm{Gpc}/h)^3$ and resolving scalar modes
  down to $k_{max}=168h/\textrm{Mpc}$.  The dashed, nearly vertical, line is
  the reconstruction noise power. The non-zero slope of the solid
  line is the deviation from scale-free.}
\end{figure}

Similar arguments are used to find the achievable error on the
spectral index $n_t$.  Properly considering the degeneracy with
$r$, the error on $n_t$ is:
\begin{equation}
\Delta n_t = F
  \left[\left(\frac{2\pi}{k_{max}}\right)^{3}\frac{1}{r
  A_s V}\right]^{1/2},
\end{equation}
where $F$ is a function of the combination of parameters
$V_H/(k_{max}^3rA_s)$. In the limit that $P_h(k=aH)\gg N_h$, which
is the limit in which a measurement of $n_t$ is possible, $F$
is approximately 6.  Assuming the same volume and redshift
as above, and that $r=0.1$, the consistency relation is tested at
the 2 sigma level for
$k_{max}=168h/\textrm{Mpc}$.  The tensor power spectrum and error bars for
this scenario are shown in Fig.~\ref{f:tensorpower}.  

Such a measurement is very futuristic indeed, requiring a nearly
filled array with greater than thousand kilometre baselines. 
Note that
such an experiment would be sensitive to $r$ down to the $10^{-6}$
level. Also, higher redshifts contain even more information,
though their observation is technically more challenging.

\sectionk{Discussion}
Aside from the technical
challenge of mapping the 21\,cm signal over hundreds of cubic
gigaparsecs and down to
scales smaller than a megaparsec, there may be other competing
effects that could hinder a detection.
Of primary concern is weak
lensing which also shears observed
structures, creating apparent local anisotropies.
The weak lensing shear is
of order a few percent, and is thus many orders of magnitude greater
than gravity wave shear.  
However, the 3D map of gravity wave shear will be
transverse, transforming intrinsically as a tensor.  
To linear order, the lensing pattern is the gradient
of a scalar.  Even at higher order, lensing always maps
one point in space to another and is thus at most vector like.
This test does not exist for the CMB or lensing due to the lower
dimensionality of these probes.

Also of concern is the preservation of the anisotropy on
small scales.
The scale corresponding to $k=168h/\textrm{Mpc}$ is still 
larger than the Jeans length at these redshifts, and
as such hydrogen should trace the dark matter.  However, 
the evolution of scalar perturbations is mildly
nonlinear, and it is possible that this evolution will erase the
anisotropy.  Detailed analysis of the nonlinear erasure of
the anisotropy is deferred to future investigation.

There has been much recent interest in searching for
anisotropy, and this has some implications for the fossil signal.  The
constraints on quadrupolar isotropy in LSS by \citet{arXiv:1003.0673}
should already imply a weak constraint
at the $r\lesssim10^6$ level.  Constraints from the CMB are
not relevant however, since modes spanning the
surface of last scatter remain super horizon today.

CMB B-modes will be the most sensitive probe of
primordial gravity waves in the next generation of experiments. 
However, fossils may eventually be sensitive well
below the limits of the CMB.

\begin{acknowledgments}

We would like to thank Patrick McDonald, Latham Boyle, Adrian Erickcek,
Neil Barnaby, Neal Dalal, Chris Hirata and Eiichiro Komatsu for
helpful discussions.
KM is supported by NSERC Canada.

\end{acknowledgments}

\bibliography{spires,externalrefs}

\end{document}